\documentstyle[epsfig,12pt]{article}

\begin{document}

\vspace*{-15mm}

\begin{center}
  {\Large {\bf Topologically Massive Gauge Theory: \\
               A Lorentzian Solution}} \\[5mm] 
  {\large K. Saygili\footnote{Electronic address: ksaygili@yeditepe.edu.tr}} 
  \\[3mm]
  {Department of Mathematics, Yeditepe University,} \\ 
  {Kayisdagi, 34755 Istanbul, Turkey} \\[10mm]

  Abstract: \\

\end{center}

  We obtain a lorentzian solution for the topologically massive 
non-abelian gauge theory on AdS space $\tilde{H}^{3}$ by means of 
a $SU(1, 1)$ gauge transformation of the previously found abelian 
solution. There exists a natural scale of length which is determined 
by the inverse topological mass $\nu \sim ng^{2}$. In the topologically 
massive electrodynamics the field strength locally determines the gauge 
potential up to a closed $1$-form via the (anti-)self-duality equation. We 
introduce a transformation of the gauge potential using the dual field 
strength which can be identified with an abelian gauge transformation. Then 
we present the map $\pi :\tilde{H}^{3} \longrightarrow \tilde{H}^{2}_{+}$ 
including the topological mass which is the lorentzian analog of the Hopf 
map. This map yields a global decomposition of $\tilde{H}^{3}$ as a trivial 
$\tilde{S}^{1}$ bundle over the upper portion of the pseudo-sphere 
$\tilde{H}^{2}_{+}$ which is the Hyperboloid model for the Lobachevski 
geometry. This leads to a reduction of the abelian field equation onto 
$\tilde{H}^{2}_{+}$ using a global section of the solution on $\tilde{H}^{3}$. 
Then we discuss the integration of the field equation using the Archimedes 
map ${\mathcal{A}}:\tilde{H}^{2}_{+}-\{N\} \longrightarrow \tilde{C}^{2}_{P}$. 
We also present a brief discussion of the holonomy of the gauge potential and 
the dual field strength on $\tilde{H}^{2}_{+}$.

\newpage

\section{Introduction}

   Topologically massive gravity and gauge theory are dynamical
theories which are specific to three dimensions \cite{DJTS1, DJTS2},
\cite{DJTS3}. They are qualitatively different from Einstein gravity 
and Yang-Mills gauge theory beside their mathematical elegance and 
consistency. 

   The most distinctive feature of the topologically massive gauge 
theories is the existence of a natural scale of length which is introduced 
by the topological mass: $[\nu]=[g]^{2}=L^{-1}$ \cite{D}, \cite{J}, 
in the geometric units ($h=1$, $c=1$). The euclidean Dirac ``monopole'' 
type solution of the Maxwell-Chern-Simons (MCS) electrodynamics on the 
de Sitter (dS) space $\tilde{S}^{3}$ is an example of the essential new 
feature introduced by the topological mass \cite{ANS, Saygili}. 
This is a physical system which intrinsically possesses the features of 
both gauge theory and gravity \cite{ANS}. 

   This solution can be embedded into the Yang-Mills-Chern-Simons 
(Y\-M\-C\-S) gauge theory by means of a $SU(2)$ gauge transformation 
\cite{Saygili}. The Hopf map 
$\pi :\tilde{S}^{3} \longrightarrow \tilde{S}^{2}$ including the 
topological mass $\nu=ng^{2}$ yields a decomposition of $\tilde{S}^{3}$ 
as a non-trivial $\tilde{S}^{1}$ bundle over $\tilde{S}^{2}$. The 
stereographic projection 
${\mathcal{S}}:\tilde{S}^{3} \longrightarrow \mathbf{R^{3}}$ 
provides a geometric view of the $\tilde{S}^{1}$ fibres in the Hopf map.
This leads to a reduction of the abelian field equation onto $\tilde{S}^{2}$ 
using local sections of $\tilde{S}^{3}$ \cite{Saygili}. In the topologically 
massive electrodynamics the field strength locally determines the gauge 
potential up to a closed $1$-form \cite{Saygili} via the self-duality 
equation \cite{TPN}, \cite{Deser1}. A Wu-Yang type construction \cite{WuYang2},
\cite{Ryder} leads to a gauge function which can be expressed in terms of the 
magnetic or the electric charges \cite{Saygili}. In geometrical terms, the 
quantization of the topological mass reduces to the quantization of the 
inverse of the natural scale of length $L=2\pi\frac{1}{\nu}$ in units of the 
inverse of the fundamental length scale $\lambda=2\pi\frac{1}{g^{2}}$. The 
fundamental scale $\lambda$ is  the least common multiple of intervals over 
which the gauge function is single-valued and periodic for any integer $n$ 
\cite{Saygili}. The integral of the field equation reduces to the formula 
for the area of a rectangle \cite{Saygili} using the Archimedes map 
${\mathcal{A}}:\tilde{S}^{2}-\{N, S\} \longrightarrow \tilde{C}^{2}$ 
\cite{MdS, AF}. The geometric phase suffered by a vector upon a parallel 
transport on $\tilde{S}^{2}$  \cite{Berry, Simon}, \cite{Mostafazadeh} can 
be expressed in terms of the holonomy of the gauge potential or the dual-field 
strength \cite{Saygili}.

  We present an analogous discussion of the lorentzian case by replacing 
the group $SU(2)$ with $SU(1, 1)$ \cite{Bargmann}, \cite{Gilmore} and the 
dS space $\tilde{S}^{3}$ with anti-de Sitter (AdS) $\tilde{H}^{3}$ 
\cite{Sharpe}. We embed the abelian solution on the AdS space $\tilde{H}^{3}$ 
which is given in \cite{ANS} into the YMCS theory by means of a $SU(1, 1)$ 
gauge transformation.

   Then we return to the abelian case. The discussion of (anti-)self-duality 
follows the same line of reasoning \cite{Saygili} for the self-duality 
\cite{TPN}, \cite{Deser1}. We introduce a transformation of the gauge 
potential using the dual field strength \cite{Saygili}. This transformation 
can be identified with an abelian gauge transformation.

  We present the map $\pi : \tilde{H}^{3} \longrightarrow \tilde{H}^{2}_{+}$ 
including the topological mass. This is the lorentzian analogue of 
the Hopf map in the euclidean case \cite{Saygili}. This yields a global 
decomposition of the AdS space $\tilde{H}^{3}$ as a trivial $\tilde{S}^{1}$ 
bundle over the upper portion of the pseudo-sphere $\tilde{H}^{2}_{+}$
\cite{AzcarragaIzquierdo}. This leads to a reduction of the field equation 
onto $\tilde{H}^{2}_{+}$ using a global section of the solution on 
$\tilde{H}^{3}$. The pseudo-sphere $\tilde{H}^{2}_{+}$ is the Hyperboloid 
(Minkowski) model of the Lobachevski geometry \cite{DFN, Kuhnel}. This can 
be mapped to the Poincare disc $\tilde{D}^{2}_{P}$ of radius 
$r=\frac{1}{\nu}$ by a stereographic projection \cite{DFN}. 

  We discuss the integration of the field equation 
on $\tilde{H}^{2}_{+}$ using the lorentzian Arc\-hi\-me\-des map 
${\mathcal{A}}:\tilde{H}^{2}_{+}-\{N\} \longrightarrow \tilde{C}^{2}_{P}$ 
from the pseudo-sphere $\tilde{H}^{2}_{+}-\{N\}$ to the cylinder 
$\tilde{C}^{2}_{P}=R \times \tilde{S}^{1}_{P}$ where $\tilde{S}^{1}_{P}$ 
is the ideal circle enclosing the Poincare disc $\tilde{D}^{2}_{P}$. 
We also present a brief discussion of the holonomy \cite{Berry, Simon}, 
\cite{Mostafazadeh} of the gauge potential and the dual-field strength 
on $\tilde{H}^{2}_{+}$.

  The topological mass $\nu \sim ng^{2}$ is not quantized in the present 
discussion.

\section{The Non-Abelian Gauge Theory}

\subsection{Yang-Mills-Chern-Simons Theory}

  The topologically massive YMCS theory is given by the 
dimensionless action

\begin{eqnarray} \label{YMCSaction}
S_{YMCS} &=& S_{YM}+S_{CS} \\
&=& -\frac{1}{2} \left[ \int tr \bigg( F \wedge *F \bigg) 
+ \nu \int tr \left( F \wedge A 
+ \frac{1}{3} g A \wedge A \wedge A \right) \right] , \nonumber 
\end{eqnarray}

\noindent where $\nu$ is the topological mass. We include the factor, 
in the action (\ref{YMCSaction}), containing the gauge coupling constant 
$g$ in the expressions for the field. Because in our solution the underlying 
geometry forces the introduction of the gauge coupling constant into the 
formulas. This provides the strength $g$ for the potentials upon assuming 
quantization of the topological mass: $\nu \sim ng^{2}$. Note that the sign 
of the topological mass in the action (\ref{YMCSaction}) is opposite to the
euclidean case \cite{Saygili} because of the conventions. The YMCS 
action (\ref{YMCSaction}) yields the field equation 
 
\begin{eqnarray} \label{YMCSfieldequation}
D*F + \nu F = 0 ,
\end{eqnarray}

\noindent  where $D$ is the gauge covariant exterior derivative. 
The field $2$-form also satisfies the Bianchi identity
 
\begin{eqnarray} \label{Bianchi}
DF=0 .
\end{eqnarray}

\noindent 

  The action (\ref{YMCSaction}) is not invariant under non-abelian 
gauge transformations

\begin{eqnarray} \label{gaugetransformation}
A'=U^{-1}AU-\frac{1}{g}U^{-1}dU .
\end{eqnarray}

\noindent It changes by 

\begin{eqnarray} \label{Windingnumber}
W=-8\pi^{2}\frac{\nu}{g^{2}} \, w ,
\end{eqnarray}

\noindent neglecting a surface term that vanishes under suitable 
asymptotic convergence conditions on $U$ \cite{DJTS1, DJTS2}, 
\cite{DJTS3}. Here the expression $w$ which is given as

\begin{eqnarray} \label{windingnumber}
w=\frac{1}{48\pi^{2}} 
\int tr \left( U^{-1}dU \wedge U^{-1}dU \wedge U^{-1}dU \right) ,
\end{eqnarray}

\noindent

\noindent corresponds to the winding number of the gauge transformation 
\cite{DJTS1, DJTS2}, \cite{DJTS3}, if the gauge group is compact, 
for example $SU(2)$. In this case the large gauge transformations, 
which are labeled by the winding number $w$, have a non-trivial 
contribution to the action.

    In the lorentzian case if we demand the expression $\exp(iS)$ 
to be gauge invariant in order to have a well-defined quantum theory 
via path-integrals \cite{DJTS1, DJTS2}, \cite{DJTS3}, then the change 
(\ref{Windingnumber}) can be tolerated if the topological mass 
is quantized as

\begin{eqnarray} \label{quantizationoftopologicalmass}
\nu=\frac{1}{4\pi} ng^{2} . 
\end{eqnarray}

\noindent We shall refer the relation: $\nu \sim ng^{2}$ as the quantization 
of the topological mass no matter if $n$ is an integer or not. Because 
this relation will naturally arise in our solutions. Further, note that
the group $SU(1, 1)$ is not compact.

\subsection{The Natural Scale of Length}

  We shall consider the YMCS theory over a spacetime with a co-frame 
consisting of the ``modified'' left-invariant basis $1$-forms of 
Bianchi type $VIII$ in Euler parameters \cite{ANS}. The topological 
mass introduces a geometric scale of length. We shall use an intrinsic 
arclength parameterization where the arclength parameters are independent 
of the length scale determined by the inverse topological mass.

  We scale the unmodified co-frame with the dimensionful factor 
$\frac{1}{\nu}$. This yields 

\begin{eqnarray} \label{coframe}
\omega^{1} &=& -\cos(\nu\psi)d\theta-\sin(\nu\psi)\sinh(\nu\theta)d\phi
\nonumber  , \\
\omega^{2} &=& -\sinh(\nu\psi)d\theta+\cos(\nu\psi)\sinh(\nu\theta)d\phi , \\
\omega^{3} &=& d\psi+\cosh(\nu\theta)d\phi ,\nonumber 
\end{eqnarray}

\noindent in terms of the intrinsic (half) arclength parameters

\begin{eqnarray} \label{changeofvariablesH3}
& & \theta=\frac{1}{2}R\tilde{\theta} \hspace*{5mm} , \hspace*{5mm}
\phi=\frac{1}{2}R\tilde{\phi} \hspace*{5mm} , \hspace*{5mm}
\psi=\frac{1}{2}R\tilde{\psi} , \\
& & \hspace*{3mm} =\frac{1}{\nu} \tilde{\theta} \hspace*{17mm} 
=\frac{1}{\nu} \tilde{\phi} \hspace*{18mm}
=\frac{1}{\nu} \tilde{\psi} \nonumber 
\end{eqnarray}

\noindent which have the dimension of length: $[\theta]=[\phi]=[\psi]=L$. 
The modification given in equation (\ref{coframe}) amounts to scaling 
the Cartan-Killing metric by the factor $\frac{1}{\nu^{2}}$

\begin{eqnarray} \label{metricH3}
& & ds^{2} = \eta_{ab}\omega^{a}\omega^{b}  \hspace*{10mm} 
\eta_{ab}=diag(-1, -1, 1) \\
& & \hspace*{7mm} = -d\theta^{2}+d\phi^{2}+2\cosh(\nu\theta)d\phi d\psi
+d\psi^{2} \nonumber \\
& & \hspace*{7mm} =-[d\theta^{2}
+\sinh^{2} \left ( \nu\theta \right ) d\phi^{2}]
+[d\psi+\cosh(\nu\theta)d\phi]^{2} , \nonumber 
\end{eqnarray}

\noindent which yields the AdS space $\tilde{H}^{3}$. The 
\textit{radius} of the AdS space is also scaled by the same factor: 
$R=\frac{1}{\nu}\tilde{R}=\frac{1}{\nu}2$. The parameters $\theta$, 
$\phi$, $\psi$ (\ref{changeofvariablesH3}) respectively represent the 
half-length of arcs which correspond to the Eulerian parameters 
$\tilde{\theta}=\nu\theta$, $\tilde{\phi}=\nu\phi$, $\tilde{\psi}=\nu\psi$ 
on the AdS space $\tilde{H}^{3}$ of \textit{radius} $R=\frac{2}{\nu}$. Thus 
(\ref{metricH3}) is the metric on the AdS space $\tilde{H}^{3}$ of 
\textit{radius} $R=\frac{2}{\nu}$ which is parameterized in terms of the 
(half) Eulerian arclengths. The scalar curvature $\mathcal{R}$ of this space 
is determined by its \textit{radius} $R$ 

\begin{eqnarray} \label{cosmologicalconstant}
{\mathcal{R}}=-\frac{6}{R^{2}}=-\frac{3}{2} \,\, \nu^{2} .
\end{eqnarray}

\noindent This can also be verified from the metric (\ref{metricH3}). 
Note that the arclength parameters are independent of the length scale 
which is determined by the inverse topological mass. We shall only 
consider the degrees of freedom which are associated with the intrinsic 
arclengths. 

   The Maurer-Cartan equations 
$d\omega^{i}=\frac{1}{2}C^{i}_{jk} \omega^{j}\wedge\omega^{k}$
for the co-frame (\ref{coframe}) yields

\begin{eqnarray}\label{MaurerCartan}
d\omega^{1}=\nu \omega^{2}\wedge\omega^{3}
\hspace*{3mm} , \hspace*{3mm}
d\omega^{2}=\nu \omega^{3}\wedge\omega^{1}
\hspace*{3mm} , \hspace*{3mm}
d\omega^{3}=-\nu \omega^{1}\wedge\omega^{2} .
\end{eqnarray}

\noindent The co-frame (\ref{coframe}) determines a unique orientation 
on the AdS space $\tilde{H}^{3}$ and these satisfy the Hodge duality 
relations

\begin{eqnarray}\label{Hodge}
*\omega^{1}=-\omega^{2}\wedge\omega^{3} 
\hspace*{3mm} , \hspace*{3mm}
*\omega^{2}=-\omega^{3}\wedge\omega^{1} 
\hspace*{3mm} , \hspace*{3mm}
*\omega^{3}=\omega^{1}\wedge\omega^{2} .
\end{eqnarray}

\noindent The Maurer-Cartan equations (\ref{MaurerCartan}) and the 
Hodge-duality relations (\ref{Hodge}) for the basis (\ref{coframe}) 
immediately lead to the result that the MCS field equation: 
$d(*F+\nu A)=0$ will be identically satisfied for the gauge potential 
$1$-form: $A=- \frac{\nu}{g} \, \omega^{3}$ \cite{ANS}. The field 
equation is given by the derivative of the (anti-)self-duality 
condition: $*F+\nu A=0$ for the topologically massive abelian 
gauge fields \cite{TPN}, \cite{Deser1}. The MCS action vanishes 
for this potential: $S_{MCS}[A]=0$. 

   This AdS space $\tilde{H}^{3}$ can be embedded into 
the space $\mathbf{R^{4}}$ with signature $(+, +, -, -)$

\begin{eqnarray} \label{embeddingS3}
(y^{1})^{2}+(y^{2})^{2}-(y^{3})^{2}-(y^{4})^{2}=R^{2} 
\hspace*{3mm} ,  \hspace*{3mm}
R=\frac{2}{\nu}, 
\end{eqnarray}

\noindent by the correspondence

\begin{eqnarray} \label{correspondenceS3}
& & y^{1}= R \cosh \left ( \nu\frac{\theta}{2} \right ) 
\cos \left ( \nu\frac{\psi+\phi}{2} \right ) 
\hspace*{3mm} , \hspace*{3mm}
y^{2}= R \cosh \left ( \nu\frac{\theta}{2} \right ) 
\sin \left ( \nu\frac{\psi+\phi}{2} \right ) , \nonumber \\
\\
& & y^{3}= R \sinh \left ( \nu\frac{\theta}{2} \right ) 
\cos \left ( \nu\frac{\psi-\phi}{2} \right ) 
\hspace*{3mm} , \hspace*{3mm}
y^{4}= R \sinh \left ( \nu\frac{\theta}{2} \right ) 
\sin \left ( \nu\frac{\psi-\phi}{2} \right ) , \nonumber 
\end{eqnarray}

\noindent where $R=\frac{2}{\nu}$. The flat metric 

\begin{eqnarray} \label{flatR4metric}
ds^{2}= (dy^{1})^{2}+(dy^{2})^{2}-(dy^{3})^{2}-(dy^{4})^{2} ,
\end{eqnarray}

\noindent on $\mathbf{R^{4}}$ reduces to (\ref{metricH3}) 
with this correspondence. 

  We shall define the map 
$\pi :\tilde{H}^{3} \longrightarrow \tilde{H}^{2}_{+}$ 
including the topological mass in section 3.2. We also scale the 
\textit{radius} of the unit hyperboloid of $2$-sheets $H^{2}$
(in euclidean $\mathbf{R^{3}}$) by the same factor $\frac{1}{\nu}$

\begin{eqnarray} \label{embeddingS2}
(x^{1})^{2}+(x^{2})^{2}-(x^{3})^{2}=-r^{2} 
\hspace*{3mm} ,  \hspace*{3mm}
r=\frac{1}{\nu} . 
\end{eqnarray}

\noindent The correspondence with the $\tilde{H}^{2}$ 
metric is given by

\begin{eqnarray} \label{correspondenceS2}
x^{1}= r \sinh (\nu\theta) \cos (\nu\phi) 
\hspace*{3mm} , \hspace*{3mm}
x^{2}= r \sinh (\nu\theta) \sin (\nu\phi) 
\hspace*{3mm} , \hspace*{3mm}
x^{3}= r \cosh (\nu\theta) ,  
\end{eqnarray}

\noindent where $r=\frac{1}{\nu}$. This provides an embedding of the 
upper portion of the hyperboloid $\tilde{H}^{2}_{+}$ with \textit{radius} 
$r=\frac{1}{\nu}$ into the space $\mathbf{R^{3}}$ with signature 
$(+, +, -)$. The flat metric  

\begin{eqnarray} \label{flatR3metric}
ds^{2}=(dx^{1})^{2}+(dx^{2})^{2}-(dx^{3})^{2} , 
\end{eqnarray}

\noindent on $\mathbf{R^{3}}$ reduces to

\begin{eqnarray} \label{defectedS2metric}
ds^{2}= d\theta^{2}
+\sinh^{2} \left ( \nu\theta \right ) d\phi^{2} ,
\end{eqnarray}

\noindent on $\tilde{H}^{2}_{+}$ with this correspondence. Thus 
(\ref{defectedS2metric}) is the metric on the pseudo-sphere 
$\tilde{H}^{2}_{+}$ of \textit{radius} $r=\frac{1}{\nu}$ which is 
parameterized in terms of the length of arcs corresponding to the 
parameters $\tilde{\theta}=\nu\theta$, $\tilde{\phi}=\nu\phi$. The 
$(-)$ factor for the metric (\ref{defectedS2metric}) in (\ref{metricH3}) 
is due to our conventions. Note that the pseudo-sphere $\tilde{H}^{2}_{+}$ 
is a space-like surface \cite{DFN, Kuhnel}. 

\begin{figure}[!hb]
\begin{center}
\includegraphics[scale=0.5]{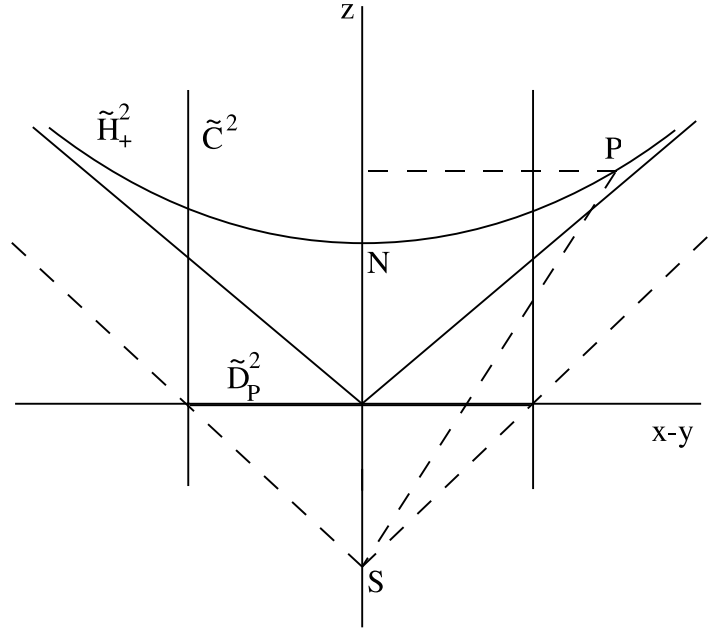}
\end{center}
\caption{The Stereographic projection 
${\mathcal{S}}:\tilde{H}^{2}_{+} \longrightarrow \tilde{D}^{2}_{P}$
and The Archimedes map 
${\mathcal{A}}:\tilde{H}^{2}_{+} \longrightarrow \tilde{C}^{2}$}  
\end{figure}

   This is the Hyperboloid (Minkowski) model of the Lobachevski geometry 
\cite{DFN}. The pseudo-sphere $\tilde{H}^{2}_{+}$ can be mapped to the 
Poincare disc with radius $r$ by a stereographic projection through the point 
$S(0, 0, -r)$ as shown in Figure 1, \cite{DFN}. The stereograpic projection 
${\mathcal{S}}:\tilde{H}^{2}_{+} \longrightarrow \tilde{D}^{2}_{P}$ is given 
as

\begin{eqnarray}\label{stereographicprojection}
& & {\mathcal{S}}(x^{1}, x^{2}, x^{3})
=\left (u^{1}=\frac{r}{r+x^{3}} x^{1}, \,\, 
u^{2}=-\frac{r}{r+x^{3}} x^{2} \right ) 
\\
& & \hspace*{24mm} 
=\left ( u^{1}=r'\cos(\nu\phi), \,\,  u^{2}=-r'\sin(\nu\phi) \right) , 
\nonumber \\
& & \hspace*{26mm} 
w=u^{1}+iu^{2}=r'\exp(-i\nu\phi) \hspace*{5mm} , \hspace*{5mm} 
r'=r\tanh \left ( \frac{\nu\theta}{2} \right ) . \nonumber
\end{eqnarray}

\noindent Note that $|w|=r' < r$. The metric (\ref{defectedS2metric}) becomes

\begin{eqnarray} \label{stereographicdefectedS2metric}
& & ds^{2}= \frac{4r^{4}}
{ \left ( r^{2}- |w|^{2} \right ) ^{2}}
|dw|^{2} \\
& & \hspace*{7mm} = \frac{4r^{4}}
{ \left [ r^{2}- (u^{1})^{2}-(u^{2})^{2} \right ] ^{2}}
\left [ (du^{1})^{2}+(du^{2})^{2} \right ] , \nonumber
\end{eqnarray}

\noindent on $\tilde{D}^{2}_{P}$. More precisely, the parameter $\theta$ is 
the length of an arc on the hyperbola which is determined by the intersection 
of the pseudo-sphere with a plane containing the $z$-axis. This becomes the 
hyperbolic radial arclength in $\tilde{D}^{2}_{P}$. The parameter $\phi$ is 
the length of an arc on the ideal circle $\tilde{S}^{1}_{P}$ of radius $r$ 
which encloses the Poincare disc.

  The range of the parameters $\theta$, $\phi$, $\psi$ are determined 
by the topological mass. The arclength $\phi$ uniquely defines the angle 
$\tilde{\phi}=\nu\phi$ for any parallel on $\tilde{H}^{2}_{+}$. 
The length of an arc, on the parallel determined by 
$\theta=\frac{1}{\nu}\tilde{\theta}$, which corresponds to 
the angle $\tilde{\phi}$ is

\begin{eqnarray}
\frac{\sinh (\nu\theta)}{\nu} \tilde{\phi}=\sinh (\nu\theta)\phi .
\end{eqnarray}

\noindent This reduces to the arclength $\phi=\frac{1}{\nu} \tilde{\phi}$
on the \textit{equator}: $\cosh(\nu\theta)=\sqrt{2}$. The coordinate $\phi$ 
which is well-defined for any point except the north pole $N(0, 0, r)$ on 
$\tilde{H}^{2}_{+}$ changes from $\phi=0$ to $\phi=2\pi\frac{1}{\nu}$ round 
about any parallel. 

\subsection{The Topologically Massive Non-abelian Solution}

   First consider the abelian solution \cite{ANS} which is given 
by the gauge potential $1$-form:

\begin{eqnarray} \label{abelianpotentialS3}
& & A= -\frac{\nu}{g} \, \omega^{3} \, \tau_{3} \\
& & \hspace*{4mm} = -\frac{\nu}{g} \, 
\Big[ d\psi+ \cos(\nu\theta) d\phi \Big] \tau_{3} .
\nonumber
\end{eqnarray}
   
\noindent Here $\tau_{i}=\frac{\rho_{i}}{2i}$ where 
$\rho_{i}$ 

\begin{eqnarray} \label{SU(1,1)generators}
\rho_{1}= \left( \begin{array}{cc}
0 & 1 \\
-1 & 0 
\end{array} \right) \hspace*{3mm} , \hspace*{3mm}
\rho_{2}= \left( \begin{array}{cc}
0 & -i \\
-i & 0 
\end{array} \right) \hspace*{3mm} , \hspace*{3mm}
\rho_{3}= \left( \begin{array}{cc}
1 & 0 \\
0 & -1 
\end{array} \right) ,
\end{eqnarray}

\noindent are the generators of the group $SU(1, 1)$. Note that 
$\rho_{1}=i\sigma_{2}$, $\rho_{2}=-i\sigma_{1}$, $\rho_{3}=\sigma_{3}$
where $\sigma_{1}$,  $\sigma_{2}$, $\sigma_{3}$ are the Pauli spin matrices
\cite{Saygili}. These satisfy the commutation relations

\begin{eqnarray}\label{commutators}
\left [ \rho_{1} \, , \, \rho_{2} \right ] = -2i\rho_{3} 
\hspace*{3mm} , \hspace*{3mm}
\left [ \rho_{2} \, , \, \rho_{3} \right ] = 2i\rho_{1} 
\hspace*{3mm} , \hspace*{3mm}
\left [ \rho_{3} \, , \, \rho_{1} \right ] = 2i\rho_{2} .
\end{eqnarray}

\noindent The factor $\frac{\nu}{g}$ yields $\frac{1}{4\pi} ng$ 
for the strength of the potential $A$ (\ref{abelianpotentialS3}) 
upon quantization of the topological mass 
(\ref{quantizationoftopologicalmass}). This is analogous to Dirac's 
quantization condition $eg=n$ which leads to the strength $g$ for 
the non-abelian solution upon a gauge transformation of the Dirac's 
monopole potential \cite{WuYang2}, \cite{Ryder}. 

   A gauge transformation of the potential (\ref{abelianpotentialS3}) 
necessarily contains the topological mass inherently in the gauge function. 
Because it also appears in the potential (\ref{abelianpotentialS3}). This 
is basically due to the arclength parameterization. The gauge function which 
embeds the topologically massive abelian solution (\ref{abelianpotentialS3}) 
into the YMCS theory is an element of the group $SU(1, 1)$, \cite{Bargmann},
\cite{Gilmore}. An element of this group is given by normalizing the 
\textit{radius} $R$ of $\tilde{H}^{3}$ which is parameterized with the 
Eulerian arclengths as

\begin{eqnarray} \label{nonabeliangaugetransformation}
\hspace*{8mm}
& & U=\frac{1}{R} \left( \begin{array}{cc}
z^{1} & z^{2} \\
\bar{z}^{2} & \bar{z}^{1} 
\end{array} \right)  \\
& & \hspace*{4mm} =\exp{(-\nu\gamma\tau_{3})}
\exp{(-\nu\beta\tau_{2})}
\exp{(-\nu\alpha\tau_{3})} . \nonumber 
\end{eqnarray}

\noindent Here  $z^{1}=y^{1}+iy^{2}$, $z^{2}=y^{3}+iy^{4}$. We identify 
the Euler parameters as $\alpha=\phi$, $\beta=\theta$, $\gamma=\psi$.
Note that $U \rightarrow 1$ as $\nu \rightarrow 0$. 
 
   A gauge transformation (\ref{gaugetransformation}) with the gauge 
function (\ref{nonabeliangaugetransformation}) of the potential
(\ref{abelianpotentialS3}) yields the potential $1$-form

\begin{eqnarray} \label{nonabelianpotential1}
& & A' = U^{-1}AU - \frac{1}{g}U^{-1}dU \\
& & \hspace*{6mm} = - \frac{\nu}{g} \,
\bigg\{ \Big[ \sin(\nu\phi)\tau_{1} - \cos(\nu\phi)\tau_{2} \Big] 
d\theta \nonumber \\
& & \hspace*{10mm}+ \Big[ \cosh(\nu\theta)\cos(\nu\phi)\tau_{1}
+ \cosh(\nu\theta)\sin(\nu\phi)\tau_{2} 
- \sinh(\nu\theta)\tau_{3} \Big] \sinh(\nu\theta) d\phi \bigg\} . \nonumber
\end{eqnarray}

\noindent This gives rise to the field $2$-form

\begin{eqnarray} \label{nonabelianfield12}
& & F' = dA' - g A' \wedge A' = U^{-1}FU \\ 
& & \hspace*{5mm} = -\frac{\nu^{2}}{g}
\Big[ \sinh(\nu\theta)\cos(\nu\phi)\tau_{1}
+ \sinh(\nu\theta)\sin(\nu\phi)\tau_{2} 
+\cosh(\nu\theta)\tau_{3} \Big] \nonumber \\
& & \hspace*{83mm} \sinh(\nu\theta)  d\theta\wedge d\phi . \nonumber
\end{eqnarray}

\noindent The dual-field $1$-form covariantly transforms as: 
$*F'=U^{-1}*FU$.  The field equation (\ref{YMCSfieldequation}) 
and the Bianchi identity (\ref{Bianchi}) are identically satisfied 
since they also covariantly transform. This is the lorentzian analog
of the topologically massive euclidean solution \cite{Saygili} where 
the parameter $\tilde{\theta}$ becomes a pseudo-angle.

    We have three observations which are based on 
simple \textit{dimensional} arguments in the equations 
(\ref{abelianpotentialS3}), (\ref{nonabeliangaugetransformation}),
(\ref{nonabelianpotential1}), (\ref{nonabelianfield12}). These are 
exactly the same as those of the euclidean case \cite{Saygili}.

   i) The gauge function necessarily contains the topological 
mass as in (\ref{nonabeliangaugetransformation}). 

   ii) The strength $\frac{\nu}{g}$ of the abelian gauge 
potential (\ref{abelianpotentialS3}) is crucial for finding 
the field $2$-form (\ref{nonabelianfield12}) in the non-abelian 
case. One finds the correct expression for the field $2$-form 
with this choice.

   iii) This is associated with the quantization of the 
topological mass.

   Thus if the strength of the potential is given as
$\frac{\nu}{g}$ then we inevitably arrive at the condition 
(\ref{quantizationoftopologicalmass}), because this yields 
$\frac{\nu}{g}=\frac{1}{4\pi} ng$ as the strength of the 
potential. Conversely, if one starts with a potential 
with the strength $\frac{1}{4\pi}ng$, then one needs to 
use (\ref{quantizationoftopologicalmass}) in finding the 
non-abelian gauge potential $A'$ (\ref{nonabelianpotential1}) 
and the field $F'$ (\ref{nonabelianfield12}). Moreover the 
YMCS field equation reduces to the condition 
(\ref{quantizationoftopologicalmass}) which will be identically 
satisfied. Note that in this discussion the number $n$ is a free 
parameter. The gauge coupling strength and the electric charge 
are denoted with $g$. The factor of $\frac{1}{4\pi}$ in the 
topological mass (\ref{quantizationoftopologicalmass}) is included 
in the strength of the potential for the sake of conciseness.

   The action (\ref{YMCSaction}) for the potential 
(\ref{nonabelianpotential1}) reduces to the Yang-Mills 
term because the Chern-Simons piece vanishes. This is also 
equal to the change $W$ (\ref{Windingnumber}) in the action 
due to the gauge transformation (\ref{nonabeliangaugetransformation}).
These are given as

\begin{eqnarray}
S_{YMCS}[ A'] &=& S_{YM}[A']=W[U] \\
&=& -\infty . \nonumber 
\end{eqnarray}

\noindent The other term which arise in the action as a 
result of the gauge transformation given by the function 
(\ref{nonabeliangaugetransformation}) vanishes. The quantization 
of the topological mass also leads to 

\begin{eqnarray} \label{ScalaRg}
{\mathcal{R}} \sim n^{2}g^{4} ,
\end{eqnarray}

\noindent for the scalar curvature (\ref{cosmologicalconstant}).

\section{The Abelian Gauge Theory}

\subsection{Maxwell-Chern-Simons Theory}

   The action for the MCS theory is given as

\begin{eqnarray} \label{MCSaction}
S_{MCS} &=& S_{M}+S_{CS} \\
&=& -\frac{1}{2} \,\, \frac{1}{2\pi} \, \left( \int F \wedge *F  
+ \nu \int F \wedge A \right) . \nonumber 
\end{eqnarray}

\noindent Note that the sign of the topological mass is opposite 
to that in \cite{Saygili}. In order to preserve the other conventions 
of \cite{Saygili} we explicitely write $\frac{1}{2\pi}$ in 
(\ref{quantizationoftopologicalmass}) as an overall factor in the 
action. We also adopt a slight change of convention in the abelian 
potential: $A=- \frac{1}{2} \frac{\nu}{g} \omega^{3}$ and the topological 
mass is now given as: $\nu = ng^{2}$. We shall refer $\nu = ng^{2}$ 
as the quantization of the topological mass without addressing whether 
$n$ is an integer. The MCS action (\ref{MCSaction}) yields the field 
equation

\begin{eqnarray} \label{MCSfieldequation}
\left ( *d + \nu \right ) *F = 0.
\end{eqnarray}

\noindent The field $2$-form also satisfies the Bianchi 
identity $dF=0$.

   The discussion of (anti-)self-duality follows the same line of 
reasoning for the self-duality \cite{TPN}, \cite{Deser1} which is 
given in \cite{Saygili}. The topologically massive field $F$ locally 
determines the potential $A'$ up to a gauge term via the (anti-)self-duality
equation

\begin{eqnarray} \label{potentialfromfield}
*F + \nu A' = 0 \hspace*{5mm} , \hspace*{5mm} F=dA' ,
\end{eqnarray}

\noindent where $ A' = A - \alpha$, $d\alpha=0$. Furthermore we 
can treat the dual-field $1$-form  $*F$ as a gauge potential in the 
field equation (\ref{MCSfieldequation}) because of the symmetry under 
the interchange $*F \leftrightarrow \nu A$ of the equations 
(\ref{potentialfromfield}) and (\ref{MCSfieldequation}). Thus 
the field equation (\ref{MCSfieldequation}) reduces to finding the 
strength of the gauge potential which is determined by the field strength 
itself. The Maxwell and the Chern-Simons terms in the MCS action 
(\ref{MCSaction}) interchange under this symmetry as $F'=F$ is kept fixed. 

   Furthermore we can define another potential $\tilde{A}$ by 
the transformation

\begin{eqnarray} \label{generalizedconnectionk=1}
\tilde{A} = A - \frac{1}{\nu} *F ,
\end{eqnarray}

\noindent which is motivated by this symmety \cite{Saygili}. A variant 
of this transformation is also observed in \cite{Itzhaki}. In fact, 
with a similar reasoning, one can introduce higher order terms of type 
$(\frac{1}{\nu}*d)^{i} A$ where $i=1, 2, 3, ...$. The topologically 
massive gauge theory is related to the CS theory through such an 
expansion of field redefinition in \cite{GL}, \cite{LJSSVV, LJSVV}.
The new potential $\tilde{A}$ transforms as a connection under the 
abelian gauge transformations. We can interpret 
(\ref{generalizedconnectionk=1}) as an abelian gauge transformation  

\begin{eqnarray} \label{abeliangaugetransformation}
A'=A - \frac{i}{g}U^{-1}dU ,
\end{eqnarray}

\noindent of the potential $A$. We find that the gauge function is 
given as

\begin{eqnarray} \label{generalizedgaugetransformation}
& & U = \exp \left ( -i \frac{g}{\nu} \oint *F \right ) \\
& & \hspace*{4mm} = \exp \left[ ig \oint ( A - \alpha ) \right] ,
\nonumber
\end{eqnarray}

\noindent upon the identification

\begin{eqnarray}
-\frac{1}{\nu} \, *F=-\frac{i}{g} d \ln U ,
\end{eqnarray}

\noindent and using the (anti-)self-duality equation 
(\ref{potentialfromfield}).   

  We have noted that the Wu-Yang construction on $\tilde{S}^{2}$ 
reduces to integration of the field equation in case one local 
potential is used on the appropriate local chart of $\tilde{S}^{2}$ 
except a pole \cite{Saygili}. Because the field strength locally 
determines the potential via the self-duality equation. In the 
present case we shall see that we have a trivial bundle 
$\tilde{H}^{3}=\tilde{S}^{1}\times\tilde{H}^{2}_{+}$. Thus
there is a local gauge potential which is globally well-defined 
on $\tilde{H}^{2}_{+}$. This gauge potential is given by a gauge 
transformation from $A=0$ with the gauge function $U$ 

\begin{eqnarray}
& & U=\exp \left( -i\frac{g}{\nu}\Phi \right)=\exp \left( igQ \right) \\
& & \hspace*{4mm} =\exp \left[ ig \oint A  \right] . \nonumber
\end{eqnarray}

\noindent Here $\Phi$ and $Q$ are respectively the electric 
\textit{flux}/circulation and the magnetic flux which will be 
defined below. These are related via the integral of the field 
equation on $\tilde{H}^{2}_{+}$.

\subsection{The $U(1)$ Gauge Field on the Pseudo-sphere: $\tilde{H}^{2}_{+}$}
  
   We shall present the map 
$\pi :\tilde{H}^{3} \longrightarrow \tilde{H}^{2}_{+}$ 
including the topological mass in this section. For a brief discussion 
of this map in a different context without the topological mass see 
\cite{AABC, M}. The AdS space $\tilde{H}^{3}$ is globally given as 
a $\tilde{S}^{1}$ bundle over the base manifold $\tilde{H}^{2}_{+}$: 
$\tilde{H}^{3}=\tilde{S}^{1}\times\tilde{H}^{2}_{+}$, 
\cite{AzcarragaIzquierdo}. Thus the bundle is trivial and we have 
a global section over the whole pseudosphere: $\tilde{H}^{2}_{+}$. 
We remind that the effect of the topological mass is to introduce a 
natural scale of length. This map leads to a reduction of the field 
equation onto the pseudo-sphere $\tilde{H}^{2}_{+}$ using a global 
section of the (anti-)self-dual solution 
$A=- \frac{1}{2} \frac{\nu}{g} \omega^{3}$ on $\tilde{H}^{3}$ for 
which the action vanishes. This gives rise to a topologically massive 
potential which is well-defined on $\tilde{H}^{2}_{+}$.

  The map $\pi :\tilde{H}^{3} \longrightarrow \tilde{H}^{2}_{+}$ is defined as

\begin{eqnarray} \label{Hopfmap}
& & x^{1} = 2\frac{r}{R^{2}} \left ( y^{1}y^{3}+y^{2}y^{4} \right ) , 
\nonumber \\
& & x^{2} = 2\frac{r}{R^{2}} \left ( y^{2}y^{3}-y^{1}y^{4} \right ) , \\
& & x^{3} = \frac{r}{R^{2}} \left [ (y^{1})^{2}+ (y^{2})^{2}
+(y^{3})^{2}+(y^{4})^{2} \right ] . \nonumber 
\end{eqnarray}
 
\noindent The section of the $3$-sphere (\ref{correspondenceS3})
corresponding to $N: \psi=-\phi$, $y^{2}=0$ is given as 

\begin{eqnarray} \label{sectionnorth}
& & z^{1} =R \cosh \left( \frac{\nu\theta}{2} \right)
= R \frac{r}{\sqrt{r^{2}-|w|^{2}}} , \\
& & z^{2} = R \sinh \left( \frac{\nu\theta}{2} \right) \exp ( -i\nu\phi ) 
= R \frac{w}{\sqrt{r^{2}-|w|^{2}}} , \nonumber \\
& & \hspace*{10mm}
w=r\frac{x^{1}-ix^{2}}{r+x^{3}} = r\frac{z^{2}}{z^{1}}
=r\tanh \left( \frac{\nu\theta}{2} \right) \exp ( -i\nu\phi ) . \nonumber
\end{eqnarray}

\noindent Here $w$ is the stereographic projection coordinate 
on $\tilde{H}^{2}_{+}$ projected from the south pole $S(0, 0, -r)$ 
to the $xy$- plane (\ref{stereographicprojection}). The local section 
given in the equation (\ref{sectionnorth}) is well-defined on whole 
$\tilde{H}^{2}_{+}$. Thus this provides a global section since it can be 
extended for any value of $\theta$. This leads to a global trivialization 
of the bundle over whole $\tilde{H}^{2}_{+}$. Therefore we have a trivial 
bundle. Thus we have the global equivalence: 
$\tilde{H}^{3}=\tilde{S}^{1} \times \tilde{H}^{2}$.

   The inverse image of a point in $\tilde{H}^{2}_{+}$ with the 
stereographic coordinate $w$ is given by the equation : 
$z^{2}=\frac{1}{r} wz^{1}$ where $| z^{1} | ^{2} - | z^{2} | ^{2} = R^{2}$. 
This is the intersection of $\tilde{H}^{3}$ with a hyperplane passing through 
the origin which is defined by this equation. The inverse image in 
$\tilde{H}^{3}$ (\ref{correspondenceS3}) of a point with the coordinate 
$w$ (\ref{sectionnorth}) in $\tilde{H}^{2}_{+}$ is parameterized 
by $\exp(-\nu\psi\tau_{3})$ or simply $\exp(i\nu\frac{\psi}{2})$ 
(\ref{nonabeliangaugetransformation}) which generates a circle. 
The image of this circle is again the same point (\ref{sectionnorth}).

    The $1$-form $\frac{1}{2}\omega^{3}$ (\ref{coframe}) 
defines a connection on the AdS space $\tilde{H}^{3}$ 
which is considered as a circle $\tilde{S}^{1}$ bundle 
over the pseudosphere $\tilde{H}^{2}_{+}$. This gives rise 
to the potential $1$-form $A=-\frac{1}{2} \frac{\nu}{g} \omega^{3}$. 
The strength $-\frac{1}{2}\frac{\nu}{g}$ of the potential 
reduces to $-\frac{1}{2}ng$ in terms of the gauge coupling 
constant $g$ upon adopting the quantization of the topological 
mass: $\nu=ng^{2}$. 

   The abelian potential $A=- \frac{1}{2} \frac{\nu}{g} \omega^{3}$ 
on $\tilde{H}^{3}$ which is given as

\begin{eqnarray} \label{AbelianpotentialS3}
& & A =  - \frac{1}{2} \, \frac{\nu}{g} \, 
\Big[ d\psi+ \cosh(\nu\theta) d\phi \Big] \\
& & \hspace*{4mm} = - \frac{1}{g} \, \frac{1}{R^{2}} \, 
\left ( -y^{2}dy^{1}+y^{1}dy^{2}+y^{4}dy^{3}-y^{3}dy^{4} \right ) ,
\nonumber
\end{eqnarray}

\noindent yields the field $2$-form

\begin{eqnarray} \label{fieldstrengthS3}
& & F=dA = -\frac{1}{2} \, \frac{\nu^{2}}{g} \, 
\sinh ( \nu\theta ) d\theta \wedge d\phi \\
& & \hspace*{15mm} = - \frac{1}{g} \, \frac{2}{R^{2}} \,
\left ( dy^{1} \wedge dy^{2} - dy^{3} \wedge dy^{4} \right ) .
\nonumber
\end{eqnarray}

   There exists a globally defined potential $1$-form $A^{N}$ on 
$\tilde{H}^{2}_{+}$ such that the field $2$-form $F$ is globally 
expressed as $F=dA^{N}$. This potential $1$-form $A^{N}$ on 
$\tilde{H}^{2}_{+}$ is locally given by the projection of the specific 
section $N: \psi=-\phi$ of $A$ (\ref{AbelianpotentialS3}) on 
$\tilde{H}^{3}$ onto $\tilde{H}^{2}_{+}$ using the map 
$\pi: \tilde{H}^{3} \rightarrow \tilde{H}^{2}_{+}$

\begin{eqnarray} \label{AbelianpotentialS2}
& & A^{N} = + \, \frac{1}{2} \, \frac{\nu}{g} \, 
\Big[ 1 - \cosh(\nu\theta) \Big] d\phi
=-\frac{1}{2} \, \frac{1}{g} \, \, 
\frac{-x^{2}dx^{1}+x^{1}dx^{2}}{r(r+x^{3})} . \nonumber \\
\end{eqnarray}

\noindent  The potential $A^{N}$ is well-defined on whole 
$\tilde{H}^{2}_{+}$ since $x^{3} \geq r$. This potential yields 
the field $2$-form

\begin{eqnarray} \label{abelianfield2form}
& & F^{N}=F 
=-\frac{1}{2} \, \frac{\nu^{2}}{g} \, \sinh (\nu\theta) d\theta \wedge d\phi \\
& & \hspace*{16mm} 
=-\frac{1}{2} \, \frac{1}{g} \, \frac{1}{r^{3}} \, \left ( x^{1}dx^{2} 
\wedge dx^{3} + x^{2}dx^{3} \wedge dx^{1} 
+ x^{3}dx^{1} \wedge dx^{2} \right ) , \nonumber
\end{eqnarray}   

\noindent on $\tilde{H}^{2}_{+}$. 

  We could have considered the section N: $\psi = -\phi$ 
for the potential (\ref{abelianpotentialS3}) and the gauge 
function (\ref{nonabeliangaugetransformation}) in embedding 
the abelian solution into the YMCS theory. It is straightforward 
to check that the equations (\ref{nonabelianpotential1}), 
(\ref{nonabelianfield12}) are satisfied for this section. 

   The potential (\ref{AbelianpotentialS3}) is determined by 
the field (\ref{fieldstrengthS3}) on $\tilde{S}^{3}$ because 
of the Hodge-duality relations (\ref{Hodge}) for the basis 
(\ref{coframe}). Hence it is self-dual (\ref{potentialfromfield}). 
The extra $\alpha$ term (\ref{potentialfromfield}) for the potential 
$A^{N}$ (\ref{AbelianpotentialS2}) on $\tilde{H}^{2}_{+}$ vanishes upon 
a consistent choice of the appropriate local section N: $\phi=-\psi$ 
for the dual-field: $*F^{N}=(*F)^{N}$  while projecting $\tilde{H}^{3}$ 
onto $\tilde{H}^{2}_{+}$. Otherwise this gives rise to an extra term 
which can be made to vanish by an abelian gauge transformation with 
$U=\exp \left [ -i\nu \left ( \phi + \psi \right ) \right ]$ on 
$\tilde{H}^{3}$. The MCS field equation (\ref{MCSfieldequation})

\begin{eqnarray} \label{MCSfieldequationS2}
d*F^{N} + \nu F^{N} = 0 ,
\end{eqnarray}

\noindent and the (anti-)self-duality equation (\ref{potentialfromfield})

\begin{eqnarray} \label{potentialfromfieldS2}
*F^{N} + \nu A^{N} =0 ,
\end{eqnarray}

\noindent are satisfied for the potential $A^{N}$ 
(\ref{AbelianpotentialS2}) on $\tilde{H}^{2}_{+}$. 
 
   We remark that the choice of the local section $N:\psi=-\phi$ endows 
us with a potential as a local expression for the connection which is 
determined by the field itself via the (anti-)self-duality condition 
\cite{Saygili}.

\subsection{Integration of the Field Equation}

  The integration of the field equation (\ref{MCSfieldequationS2}) follows 
the same line of reasoning as in the euclidean case \cite{Saygili}. We can 
use geometric quantities such as area and arclength on the pseudo-sphere 
$\tilde{H}^{2}_{+}$ or the Poincare disc $\tilde{D}^{2}_{P}$.

     The integration of the field equation (\ref{MCSfieldequationS2}) 
over the local chart $U(P:\theta)$ of $\tilde{H}^{2}_{+}$ which is 
determined by $r \leq x^{3} \leq r\cosh(\nu\theta)$ yields

\begin{eqnarray} \label{fluxcharge}
\Phi^{N} + \nu Q^{N} = 0 . 
\end{eqnarray}

\noindent Here $\Phi$ and $Q$ are respectively the 
electric \textit{flux}/circulation and the magnetic 
flux through $U(P:\theta)$ which are defined as 

\begin{eqnarray} \label{Electricharge}
\Phi^{N} \equiv \int_{U(P:\theta)} d*F^{N} 
\hspace*{5mm} , \hspace*{5mm}
Q^{N} \equiv \int_{U(P:\theta)} F^{N} . 
\end{eqnarray}

\noindent These reduce to loop integrals over the boundary of 
the chart $U(P:\theta)$ which is given by the parallel $P:\theta$ 
upon using the Stokes theorem

\begin{eqnarray}  \label{ElectrichargeStokes}
\Phi^{N} = -\oint_{P:\theta} *F^{N} 
\hspace*{5mm} , \hspace*{5mm}
Q^{N}= -\oint_{P:\theta}  A^{N} .
\end{eqnarray}

\noindent The factors of $(-)$ sign are due to orientation on 
the parallel $P:\theta$ since it encircles the chart $U(P:\theta)$ 
in the left-handed sense. Thus we conclude, from the equations 
(\ref{fluxcharge}-\ref{ElectrichargeStokes}), that the magnetic 
flux $Q^{N}$ through $U(P:\theta)$ is determined by the electric 
circulation $\Phi^{N}$ on the boundary $P:\theta$. The electric
\textit{flux}/circulation and the magnetic flux associated with 
the potential $A^{N}$ are given as

\begin{eqnarray} \label{partialcharges}
\Phi^{N}=-\pi \frac{\nu}{g} \Big[ 1 - \cosh(\nu\theta) \Big] 
\hspace*{7mm} , \hspace*{7mm}
Q^{N}=\pi \frac{1}{g} \Big[ 1 - \cosh(\nu\theta) \Big].  
\end{eqnarray}

\noindent These diverge as $\theta \rightarrow \infty$.

  The cylindrical coordinates leads to an interpretation of the 
electric \textit{flux}/\-circulation and the magnetic flux in terms 
of area and arclength as in the euclidean case \cite{Saygili}. We 
can construct the Archimedes map 
${\mathcal{A}}:\tilde{H}^{2}_{+}-\{N\} \longrightarrow \tilde{C}^{2}_{P}$ 
as a perpendicular projection from the $z$-axis as shown in Figure 1. 
The cylinder $\tilde{C}^{2}_{P}=R \times \tilde{S}^{1}_{P}$ is given as 
the product of the real line $R$ and the ideal \textit{Poincare circle} 
$\tilde{S}^{1}_{P}$ of radius $r=\frac{1}{\nu}$. The image in 
$\tilde{C}^{2}_{P}$ of a point $P(x^{1}, x^{2}, x^{3})$ in 
$\tilde{H}^{2}_{+}-\{N\}$ is given as

\begin{eqnarray}
{\mathcal{A}}(x^{1}, x^{2}, x^{3})= 
\left( \frac{r}{\sqrt{(x^{3})^{2}-r^{2}}} \, x^{1}, 
\,\, \frac{r}{\sqrt{(x^{3})^{2}-r^{2}}} \, x^{2}, \,\, x^{3} \right) .
\end{eqnarray}

\noindent The Archimedes map ${\mathcal{A}}$ is locally area preserving. 
The gauge potential $1$-form and the field $2$-form are given as 

\begin{eqnarray}\label{Archimedesarea}
A^{N}=-\frac{1}{2}\frac{\nu^{2}}{g}a 
\hspace*{5mm} , \hspace*{5mm} 
F^{N}=-\frac{1}{2}\frac{\nu^{2}}{g}\sigma .
\end{eqnarray}

\noindent Here the $1$-form $a$ is the area of a thin 
strip on $\tilde{C}^{2}_{P}$ of base length $d\phi$ and height 
$h=x^{3}(\theta)-x^{3}(0)$ and $\sigma$ is the area $2$-form 
on $\tilde{C}^{2}_{P}$. The electric \textit{flux}/circulation and 
the magnetic flux (\ref{partialcharges}) are given as

\begin{eqnarray}
& & \Phi^{N}=\frac{1}{2}\frac{\nu^{3}}{g} \Sigma_{U} 
\hspace*{5mm} , \hspace*{5mm}
Q^{N}=-\frac{1}{2}\frac{\nu^{2}}{g} \Sigma_{U} , \\
& & \hspace*{7mm} =\pi\frac{\nu^{2}}{g}h_{U} \nonumber 
\end{eqnarray}

\noindent in terms of the area $\Sigma_{U}$ and height $h_{U}$ of 
the chart $U(P:\theta)$ or its image in $\tilde{C}^{2}_{P}$. Thus we 
can interpret the integrated field equation (\ref{fluxcharge}) simply 
as the area formula: $height=\frac{1}{base}area$ for a rectangle
of height $h=x^{3}(\theta)-x^{3}(0)$ and base $2\pi\frac{1}{\nu}$
up to an overall factor of $\frac{1}{2}\frac{\nu^{2}}{g}$.

\subsection{The Holonomy}

    The Berry phase \cite{Berry}, which corresponds to holonomy in 
a line bundle \cite{Simon}, for the group $SU(1, 1)$ is investigated
in various contexts \cite{Mostafazadeh}. The classical example of this 
is the geometric phase suffered by a vector upon parallel transport 
on the pseudosphere $H^{2}_{+}$. The geometric phase is given by a 
measure of the area suspended by a loop analogous to the euclidean 
case \cite{Mostafazadeh}, \cite{Saygili}. It has become easy to express 
this in terms of the holonomy of the topologically massive gauge potential 
or the dual-field. 

   Consider a vector $X$ which is tangent to $\tilde{H}^{2}_{+}$ at 
longitude $\phi=0$ and at latitude $\theta$. If we parallel transport 
this vector along the latitude $P:\theta$, it does not coincide with its 
initial direction after a complete revolution. But it suffers a phase 
$\tilde{\gamma}$ which is determined by a measure of the area $\Sigma_{U}$
suspended by the loop $P:\theta$ in $\tilde{H}^{2}_{+}$ \cite{Mostafazadeh}. 
This is given as

\begin{eqnarray} \label{Solidangle}
& & \tilde{\gamma}=\tilde{\Omega}=\frac{1}{r^{2}}\Sigma_{U} \\
& & \hspace*{12mm} =-2\pi [1-\cosh(\nu\theta)] , \nonumber 
\end{eqnarray}

\noindent where $\tilde{\Omega}=\frac{1}{r^{2}}\Sigma$ is 
the normalized area. It is straightforward to verify this 
by solving the equation for parallel transport: 
$\nabla_{(\phi)}X=0$ with the metric (\ref{defectedS2metric}) 
on $\tilde{H}^{2}_{+}$. This equation reduces to

\begin{eqnarray}
\frac{dZ}{d\phi}+i\nu\cosh(\nu\theta)Z=0.
\end{eqnarray}

\noindent We find

\begin{eqnarray}\label{solution}
Z(2\pi\frac{1}{\nu})=Z(0)\exp\{-i\oint_{P}\nu[1-\cosh(\nu\theta)]d\phi\} ,
\end{eqnarray}

\noindent for a complete revolution. Here $Z(0)$ and 
$Z(2\pi\frac{1}{\nu})$ respectively correspond to the initial 
and the final vectors $X_{i}$, $X_{f}$. The phase in (\ref{solution}) 
reduces to the normalized area in (\ref{Solidangle}) of the region which 
is subtented by the parallel $P:\theta$ via the Stokes theorem. The phase 
(\ref{Solidangle}) can be written in terms of the holonomy $\tilde{\Gamma}$ of 
the topologically massive gauge potential or the dual-field over the parallel 
$P:\theta$ as

\begin{eqnarray}
\tilde{\gamma}=-2\tilde{\Gamma} ,
\end{eqnarray}

\noindent where 

\begin{eqnarray} \label{Holonomy}
\tilde{\Gamma}=Q^{N}=-\frac{1}{\nu}\Phi^{N} , 
\end{eqnarray}

\noindent (\ref{partialcharges}) and the factor of $\frac{1}{g}$ 
is ignored. 

\section{Conclusion}

  We have considered lorentzian solutions of the MCS and the $SU(1, 1)$ 
YMCS theories on the AdS space $\tilde{H}^{3}$. We have embedded the abelian 
solution into the YMCS theory by means of a $SU(1, 1)$ gauge transformation.
The action for the abelian solution vanishes. Meanwhile the action for the 
non-abelian solution consists of only the Yang-Mills term and this is infinite.

  In the abelian case the topologically massive field locally determines 
the potential up to a closed $1$-form via the (anti-)self-duality equation. 
We have introduced a transformation of the gauge potential using the 
dual-field strength. This transformation can be identified with an 
abelian gauge transformation. The gauge function is given in terms
of the magnetic flux or the electric \textit{flux}/circulation. 

  Then we have introduced the map 
$\tilde{H}^{3} \longrightarrow \tilde{H}^{2}_{+}$ including 
the topological mass which is the lorentzian analogue of the Hopf 
map in the euclidean case. This map yields a global decomposition of 
$\tilde{H}^{3}$ as a trivial $\tilde{S}^{1}$ bundle over the pseudo-sphere 
$\tilde{H}^{2}_{+}$. This leads to a reduction of the abelian field equation 
onto $\tilde{H}^{2}_{+}$ using the global section $N: \psi=-\phi$ of the 
solution on $\tilde{H}^{3}$. This solution carry both magnetic flux and 
electric \textit{flux}/circulation. The magnetic flux $Q^{N}$ through 
a finite chart $U(P:\theta)$ of $\tilde{H}^{2}_{+}$ is determined by 
the electric circulation $\Phi^{N}$ on the boundary of this chart. 
The Archimedes map 
${\mathcal{A}}:\tilde{H}^{2}_{+}-\{N\} \longrightarrow \tilde{C}^{2}_{P}$ 
has led to a simple interpretation of this composite structure in terms 
of the area of a rectangle. We have also expressed the geometric phase 
suffered by a vector upon parallel transport on the pseudo-sphere 
$\tilde{H}^{2}_{+}$ in terms of the holonomy of the topologically massive 
gauge potential or the dual-field.

  These are analogous to the euclidean solutions \cite{ANS, Saygili} 
of the MCS and the $SU(2)$ YMCS theories on the $3$-sphere $\tilde{S}^{3}$. 
There exists a natural scale of length which is determined by the inverse 
topological mass $\nu \sim ng^{2}$. We have used an intrinsic arclength 
parameterization. The arclength parameters are taken to be independent of 
the length scale which is introduced by the topological mass. In geometrical 
terms, the quantization of the topological mass requires the quantization of 
the inverse natural scale of length in units of the inverse fundamental 
length scale $g^{2}$ as in the euclidean case \cite{Saygili}. However in 
the present discussion the parameter $n$ is free and it can assume an 
arbitrary value.

\end{document}